\documentclass{iau}
\usepackage{graphicx}

\title[IAU303] 
{Shock Structure and Shock Heating in
 the Galactic Central Molecular Zone}

\author[J\"urgen Ott, et al. ]   
{J\"urgen Ott$^{1,2}$, Michael Burton$^3$, Paul Jones$^3$, \and David S. Meier$^{2,1}$ }

\affiliation{$^1$National Radio Astronomy Observatory, P.O. Box O,
  1003 Lopezville Road, Socorro, NM 87801, USA; email: {\tt
    jott@nrao.edu}\\
$^2$Department of Physics, New Mexico Institute of Mining and
Technology, Socorro, NM 87801, USA;  email: {\tt
    dmeier@nmt.edu}\\
 $^3$School of Physics, University of New South Wales, Sydney NSW
 2052, Australia;  email: {\tt
    M.Burton@unsw.edu.au}, {\tt paulcojones@gmail.com}\\
}

\pubyear{2013}
\volume{303}  
\pagerange{119--126}
\jname{The Galactic Center: Feeding and Feedback in a Normal Galactic Nucleus }
\editors{L. Sjouwerman, C. Lang, \& J. Ott, eds.}
\begin{document}

\maketitle

\begin{abstract}
  We present maps of a large number of dense molecular gas tracers
  across the Central Molecular Zone of our Galaxy. The data were taken
  with the CSIRO/CASS Mopra telescope in Large Projects in the
  1.3\,cm, 7\,mm, and 3\,mm wavelength regime.
  Here, we focus on the brightness of the shock tracers SiO and HNCO,
  molecules that are liberated from dust grains under strong (SiO) and
  weak (HNCO) shocks.
  The shocks 
  may have occurred when the gas enters the bar regions and the shock
  differences could be due to differences in the moving cloud
  mass. Based on tracers of ionizing photons, it is unlikely that the
  morphological differences are due to selective photo-dissociation of
  the molecules. We also observe direct heating of molecular gas in
  strongly shocked zones, with a high SiO/HNCO ratios, where
  temperatures are determined from the transitions of ammonia. Strong
  shocks appear to be the most efficient heating source of molecular
  gas, apart from high energy emission emitted by the central
  supermassive black hole Sgr A* and the processes within the extreme
  star formation region Sgr B2.  \keywords{ISM: molecules --- Galaxy:
    center --- radio lines: ISM}
\end{abstract}

\firstsection 
\section{Introduction}

The Central Molecular Zone (CMZ) covers the inner $\sim 500$\,pc of
the Milky Way and contains a substantial amount of molecular gas
(several $10^{7}$\,M$_{\odot}$ e.g. \cite[Oka et al. 1998]{oka98}). Models show that
the gas may be funneled along the central Galactic bar into the CMZ
and eventually accretes on $x_2$ orbits that cover the inner $\sim
200$\,pc (e.g. \cite[Morris \& Serabyn 1996]{mor96}). Eventually the gas will be destroyed by
environmental influences, during the process of star formation, or, in
rare cases, the dense gas will be transported all the way further
toward Sgr A*, the central supermassive black hole of the Milky Way.

In this paper, we focus on material that is formed during
shocks. HNCO, for example, is a molecule that is formed on dust grains
and liberated into the gas phase by weak shocks
(\cite[Meier \& Turner 2005]{mei05}). Stronger shocks can be traced by SiO where the silicon
is provided by shock destruction of the dust grains themselves
(\cite[Martin-Pintado et al. 1997]{mar97}). Since the ISM conditions in the Galactic Center are
dominated by shock physics, we use these two molecules to trace the
environment as the gas accumulates in the CMZ. \cite{jon12} presents
maps of both molecules using the ATNF/CASS Mopra\footnote{The Mopra
  radio telescope is part of the Australia Telescope National Facility
  which is funded by the Commonwealth of Australia for operation as a
  National Facility managed by CSIRO. } telescope and \cite{ott08}
shows single dish Mopra maps of ammonia, a molecule which is used as a
thermometer to trace shock heating.

\section{The Shock Structure of Inflowing Gas}

\begin{figure}[b]
 \vspace*{-2.0 cm}
\begin{center}
 \includegraphics[width=5.5in]{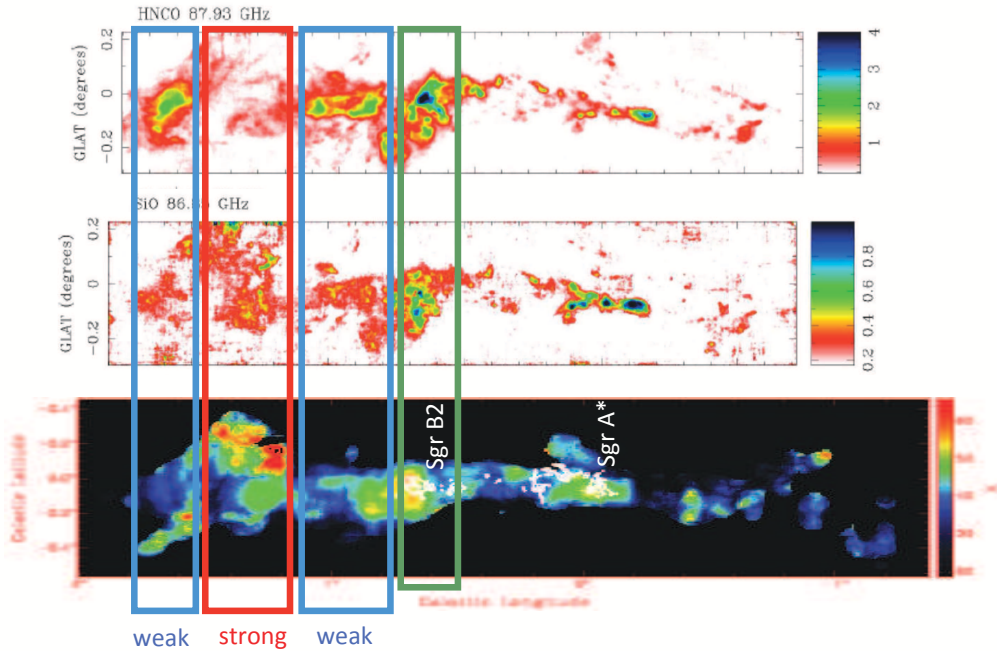} 
 \vspace*{-1.0 cm}
 \caption{Shock tracers in the CMZ: {\it Top:} HNCO, {\it Middle:}
   SiO. At the bottom we show a gas temperature map as derived from
   NH$_3$. The regions with weak shoks are marked in {\it blue}, the
   strongly shocked regions in {\it red}. The {\it green} box centers on the star
   forming region Sgr B\,2 and marks the likely point where the $x_1$
   orbits accrete on the 100\,pc ring, the location of the $x_2$
   orbits (color figure available in the online version). }
   \label{fig}
\end{center}
\end{figure}

In Fig.\,\ref{fig} we show the distribution of the HNCO
($4_{04}-3_{03}$) and SiO (2-1) line emission in the CMZ. In
particular the regions east of Sgr\,B2 show a distinct pattern of
alternating strong SiO and HNCO lines. This hints to alternating
stronger and weaker shocks, where the shocks might have occurred at
the time the gas entered the large Galactic bar or along $x_1$ orbits
as the gas moves toward the Galactic Center. Eventually the gas may
accrete on the 100\,pc ring (likely the location of the $x_2$ orbits)
that was discovered in {\it HERSCHEL} dust maps (\cite[Molinari et
al. 2011]{mol11}).  This scenario is in agreement with gas heating by
shocks. At the positions where the SiO emission dominates, the ammonia
temperature map shows much higher temperatures, compared to the places
with HNCO dominating. In addition, the velocity dispersion
increases at the same morphological features. The shock heated gas is
some of the hottest gas in the Galactic Center, outside of the
environments of compact sources.

\end{document}